\documentclass[debug, preprint, twocolumn]{rmaa}
\usepackage{natbib}

\usepackage{color}
\usepackage{./aas_macros}

\font\manual=manfnt at 7pt \def\dbend{\hbox{\raise0.9ex\hbox{\manual\char127\hspace{0.6em}}}}

\providecommand{\e}[1]{\ensuremath{\times 10^{#1}}}

\newcounter{INTERNALionstage}

\def\gtsim{\mathrel{\hbox{\rlap{\hbox{\lower4pt\hbox{$\sim$}}}\hbox{$>$}}}}
\def\lesssim{\mathrel{\hbox{\rlap{\hbox{\lower4pt\hbox{$\sim$}}}\hbox{$<$}}}}

\def\pcc{{\rm\thinspace cm^{-3}}}

\def\htwo{\mbox{{\rm H}$_2$}}

\DeclareMathAlphabet{\vib}{OML}{cmm}{m}{it}

\title{\boldmath The abundance discrepancy factor and \MakeLowercase{$t^2$} in nebulae: \\
are non-thermal electrons the culprits?}
\shorttitle{The abundance discrepancy factor, \MakeLowercase{$t^2$},
and kappa electrons}

\author{
G. J. Ferland\altaffilmark{1}, 
W.~J. Henney\altaffilmark{2},
C.~R. O'Dell\altaffilmark{3}, and
M. Peimbert\altaffilmark{4}
}
\shortauthor{Ferland et al.}

\altaffiltext{1}{Department of Physics and Astronomy, University of Kentucky, Lexington, KY 40506, USA}

\altaffiltext{2}{Centro de Radioastronom\'{\i}a y Astrof\'{\i}sica, Universidad Nacional Aut\'onoma de M\'exico, Apartado Postal 3-72, 58090 Morelia, Michoac\'an, M\'exico}

\altaffiltext{3}{Department of Physics and Astronomy, Vanderbilt University, Box 1807-B, Nashville, TN 37235}

\altaffiltext{4}{Instituto de Astronomia, Universidad Nacional Aut\'onoma de M\'exico, Apartado Postal 70-264, 04510 M\'exico D. F., M\'exico}

\fulladdresses{
\item G. J. Ferland:
Department of Physics and Astronomy, University of Kentucky, Lexington, KY 40506, USA

\item W.  J. Henney:
Centro de Radioastronom\'{\i}a y Astrof\'{\i}sica, Universidad Nacional Aut\'onoma de M\'exico, Apartado Postal 3-72, 58090 Morelia, Michoac\'an, M\'exico

\item C. R. O'Dell:
Department of Physics and Astronomy, Vanderbilt University, Box 1807-B, Nashville, TN 37235

\item M. Peimbert Instituto de Astronomia, Universidad Nacional Aut\'onoma de M\'exico, Apartado Postal 70-264, 04510 M\'exico D. F., M\'exico
}

\abstract{
Photoionization produces supra-thermal electrons,
electrons with much more energy than is found
in a thermalized gas at electron temperatures characteristic of nebulae. 
The presence of these high energy
electrons may solve the long-standing $t^2$/ADF puzzle,
the observations that abundances obtained from recombination and collisionally excited lines do
not agree, and that different temperature indicators give different results,
if they survive long enough to affect diagnostic emission lines. 
The presence of these non-Maxwellian distribution electrons are usually designated by the term kappa.
Here we use well established methods to show that the distance over which heating rates
change are much longer than the distance supra thermal electrons can travel, 
and that the timescale to thermalize these electrons are much shorter than the heating or
cooling timescales.
These estimates establish that
supra thermal electrons will have disappeared into the Maxwellian velocity distribution 
long before they affect the collisionally excited forbidden and recombination lines 
that are used for deriving abundances relative to hydrogen.
The electron velocity distribution in nebulae should be closely thermal.
}

\resumen{%
La fotoionizaci\'{o}n produce electrones suprat\'{e}rmicos, electrones mucho mas energ\'{e}ticos que
los que se encuentran en un gas termalizado con una temperatura electr\'{o}nica caracter\'{i}stica
de nebulosas gaseosas. La presencia de estos electrones de alta energ\'{i}a podr\'{i}a resolver: a) la
discrepancia entre las abundancias obtenidas a partir de l\'{i}neas prohibidas y las obtenidas a
partir de l\'{i}neas de recombinaci\'{o}n (problema $t^2$/ADF), y b) que los diferentes indicadores
de temperatura den valores diferentes. Esto ser\'{i}a posible si los electrones suprat\'{e}rmicos
sobreviven el tiempo suficiente para afectar las l\'{i}neas de emisi\'{o}n. Usamos m\'{e}todos bien
establecidos para mostrar que las distancias  en que las tasas de calentamiento cambian
son mucho mayores que las distancias que pueden viajar los electrones suprat\'{e}rmicos, y
que las  escalas de tiempo para termalizar a estos electrones son mucho menores que las
escalas de calentamiento o enfriamiento. Estas estimaciones establecen que los electrones
suprat\'{e}rmicos se maxwellianizan mucho antes de que puedan afectar a las l\'{i}neas prohibidas
colisionalmente excitadas y a las l\'{i}neas de recombinaci\'{o}n que se usan para obtener las
abundancias relativas. La distribuci\'{o}n  electr\'{o}nica de velocidades en las nebulosas debe
ser muy cercana a la Maxwelliana.
}

\addkeyword{atomic processes}
\addkeyword{galaxies: active}
\addkeyword{methods: numerical}
\addkeyword{molecular processes}
\addkeyword{radiation mechanisms}

\begin{document}

\maketitle
\clearpage

\section{Introduction}

The electron kinetic temperature is one of the most fundamental characteristics in a photoionized
nebula such as an H II region or a planetary nebula. 
These temperatures are derived from ratios of emission lines in the optical spectrum 
and this methodology has been  
a major tool in the analysis of the nebular spectra for 60 years
\citep{2002RMxAC..12....1O}.
These temperatures are usually about 10,000 Kelvin and are called the
electron kinetic temperatures ($T\rm_{e}$) since they reflect the energy distribution of the free electrons in the gas.
$T\rm_{e}$ is different from other temperatures encountered in the treatment of a nebula,
e.g. the ionizing star temperature, the ionization temperature, and the excitation temperature 
derived from the ratio of populations in atomic levels.
If one knows $T\rm_{e}$, one can then use other emission lines to determine
the relative abundance of different atoms, a process reviewed in \citet{AGN3}, hereafter AGN3.

Collisionally-excited forbidden lines of the heavy elements are often  the strongest lines in a nebula's spectrum.
Their emissivity increases sharply with higher $T\rm_{e}$. This is in contrast with lines produced during the recombination of electrons with ions, where the emissivity increases with lower $T\rm_{e}$.
While hydrogen and helium recombination lines are strong, 
recombination lines of the heavy elements are usually weak in the observed spectrum because their strength approximately scales with their relative abundance. 
Because of the different temperature dependencies, abundances determined from ratios of
recombination lines will be far less temperature sensitive than those determined from the 
ratio of collisionally excited and recombination lines.
The state of the art is that  faint forbidden and recombination lines from heavy elements can now be measured 
and abundances from the two methods compared
\citep{2003ApJ...584..735P,2004MNRAS.355..229E}.

There is a long-standing riddle when applying $T\rm_{e}$ for the determination of  abundances relative to hydrogen.
One generally derives higher relative abundances from recombination lines than from the forbidden 
lines. This is called the ``abundance discrepancy factor'' (ADF) problem,
and a possible origin lies in temperature fluctuations, parameterized as ``$t^2$''  
\citep{Peimbert1967,1993ApJ...414..626P}. 
The presence of temperature fluctuations causes the abundances determined from the
forbidden lines to underestimate the abundances, accounting for the ADF.

The ADF or $t^2$ problem is largest in planetary nebulae, 
with an ADF sometimes more than an order of magnitude
\citep{2000MNRAS.312..585L}.
It is smaller in H II regions, such as the Orion nebula, 
but the ADF is still nearly a factor of two \citep{2013ApJ...778...89P}. 
The $t^2$ values predicted by photoionization models with constant density are in the 
0.002 - 0.02 range with typical values around 0.004,
while in many cases the $t^2$ observed values are higher, in the 0.02 to 0.05 range.
Large temperature fluctuations could be caused by many processes including a) density fluctuations,
b) shocks, c) shadowed regions, d) chemical composition inhomogenieties, 
or e) variations in the fluxes from the ionizing stars.
These fluctuations are caused by real changes in temperature caused by different regions
contributing along a line of sight and within the spatial resolution of the observations, 
with each region having a
well defined electron kinetic temperature corresponding to 
a Maxwellian velocity distribution.

One suggestion to account for $t^2$, discussed in three recent papers, is that the free
electrons in the ionized gas do not have a thermal velocity distribution \citep{KappaNicholls12,KappaNicholls13,KappaDopita13}. 
These authors propose that a significant population of supra-thermal electrons exist in the ionized gas.
They use the ``kappa'' formulation \citep{1968JGR....73.2839V} to describe the velocity distribution of these non-thermal electrons.
The assumption of kappa is attractive because such a distribution would have more high velocity electrons than a Maxwellian distribution. 
These high-velocity electrons would be more effective in collisional excitation of the metastable levels that 
produce the forbidden lines of heavy elements, but particularly those of auroral lines,
whose relative strength is critical in determining $T_e$. 
This would mean that the temperature determined assuming a Maxwellian distribution will be over-estimated; 
therefore the emissivity of the nebular forbidden lines will be over-estimated and the relative abundance of the 
heavy elements will be systematically underestimated. 
This would also explain the ADF. 

However, the  assumption that non-thermal electrons 
could survive in an ionized gas long enough to affect the spectrum  
has not been critically examined.
We know of no detailed numerical calculation that follows the evolution of a 
non-thermal electron in an ionized gas. 
This has not been done since the arguments outlined below suggest that this is not necessary. 
Below we outline the processes and timescales that characterize a photoionized gas such 
as an H II region or planetary nebula.  
These are based on the rigorous treatment of the rates of collisions and recombinations 
that apply to an  ionized gas.
These rates determine the time scale and distance over which a non-thermal 
electron distribution would disappear.
They show that the forces establishing a thermal distribution are powerful and 
far faster than those that disturb the distribution.

In Section \ref{sec:ThermalAndKappa} we describe the basic characteristics of 
a thermal gas and a kappa gas, and in Section \ref{sec:where} we identify where kappa may play a role.
Section \ref{sec:primary} describes heating and ionization processes in photoionization
while Section \ref{sec:Orion}  gives typical numbers for the photo ionization physics 
for the Orion Nebula, a characteristic H~II region. 
In Section \ref{sec:WhatIf} we discuss what would happen if there are kappa electrons, 
and we summarize our conclusions in Section \ref{sec:conclusions}.

\section{Electron Velocity Distributions in a Thermal Gas and a Kappa Gas}
\label{sec:ThermalAndKappa}

\begin{figure}[t]
\includegraphics[scale=0.43]{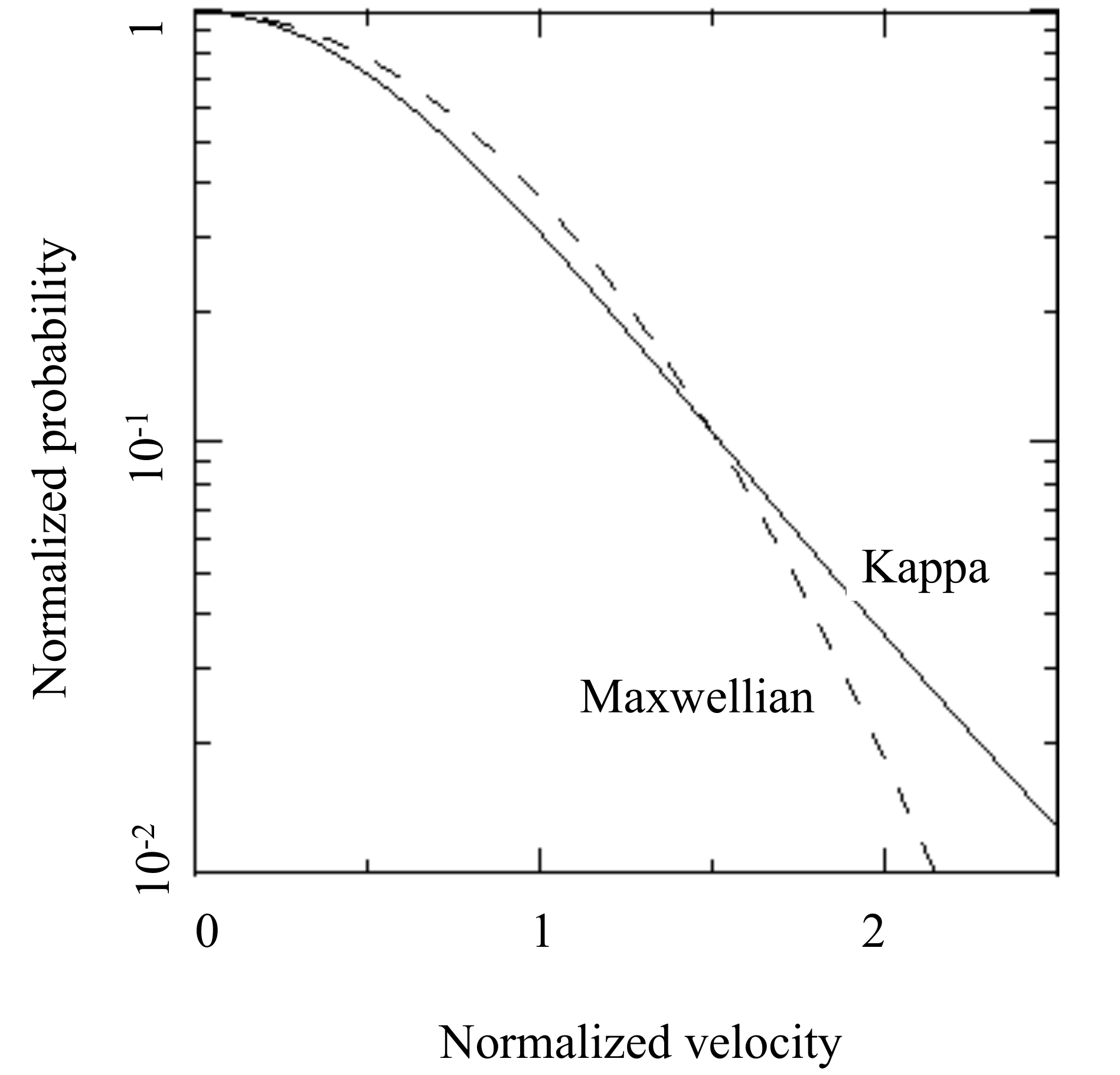}
\caption[Maxwellian vs kappa electron distributions]{Comparing a Maxwellian
with a kappa electron velocity distribution.
The kappa distribution has more high-energy electrons than a thermal gas.
Smaller values of kappa correspond to greater deviations from a Maxwellian
while a thermal distribution has a kappa of infinity.
Equation \ref{eq:t2kappa} shows the relation between $t^2$ and kappa.
\label{fig:MaxwellianKappa}}
\end{figure}

In a collisionless gas, a gas where particles do not collide with one another,
electrons will maintain whatever energy and velocity they originally have.
Collisions between particles causes them to share kinetic energy, and if a sufficient number
of collisions occur, they will reach energy equipartition.
Particles in a thermal gas have undergone a sufficient number of collisions to have reached this equilibrium
and the distribution of velocities is  given by the Maxwellian distribution function. 
This distribution has a most frequent velocity, 
and at lower velocities is proportional to the velocity 
squared while at higher velocities has an exponential tail.
Such a distribution is shown as the dashed line in Figure \ref{fig:MaxwellianKappa}.
This physics is covered in, for instance, \citet{Spitzer1962}.

This is quite different from a kappa gas, where the 
electron distribution function has an excess at high energies and velocities, as shown by
the solid line in Figure \ref{fig:MaxwellianKappa}.
This must happen to some extent since photo electrons
enter the gas at a  high velocity and then decay to lower velocity. 
After enough collisions, the signature 
of the kappa gas disappears as the history of the injected electrons is erased.

To explain typically observed ADF  values,  kappas in the 20 to 50 range  are required. 
Moreover the effect of kappa can be reproduced with temperature variations of the magnitude 
\begin{equation}
\label{eq:t2kappa}
t^2= 0.96/\kappa 
\end{equation}
\citep{2013ApJ...778...89P}. 
Values of kappa larger than 1000 produce  negligible ADF values. 

As described in the following sections, high-energy photoelectrons are continuously entering
the plasma, so to some extent a high-energy kappa tail will be present.
The key question is how important these high-energy electrons are relative to the thermal electrons.
This comes down to a question of timescales - how does the thermalization timescale, the time
required for a high-energy electron to become thermal, compare with the rate non-thermal electrons
are introduced into the gas, or for them to affect the forbidden lines?  
We address this by examining timescales for equilibrium in a photoionized gas in the following sections.


\section{Where non-thermal velocity distributions apply, and tests to determine this}
\label{sec:where}
This section outlines three regions where suprathermal electrons are known to be important,
and discusses the  tests used to diagnose this condition.

Non-thermal electron distributions are  important in active regions of the sun, 
as reviewed by \citet{2013SSRv..178..271B}.
Flares are regions where suprathermal electrons are created by the
explosive release of energy following magnetic reconnection.
These very high energy electrons interact with atoms before they have time to relax.
A \emph{timescale test} is applied to determine whether suprathermal or kappa electrons will be important.
Such tests show that flare electrons will be non-thermal.

Non-thermal particle distributions are also important in certain types of shocks, also reviewed by 
\citet{2013SSRv..178..271B}.
This is most important in low-density neutral regions where the  mean free path becomes large
compared with the dimensions in the system.  The most important effect is when cold 
neutral atoms  pass through a shock and become ionized in warm ionized regions downstream.  
These cold protons can emit
before undergoing enough collisions to become thermal.  
Details of the resulting non-Maxwellian velocity distribution, for protons,
are given by \citet{2008ApJ...682..408R}.
In this case a \emph{length test} that compares the mean free path and the dimension
for changes in the system is applied and shows that non-thermal distributions are important.

The kappa formalism is only one way of dealing with non-thermal electrons.
Most spectral simulation codes, including Cloudy \citep{CloudyReview13}, solve for a population of suprathermal electrons
and include them as a general excitation process.
Suprathermal electrons are known to be important 
when high-energy photons enter
neutral regions such as the H$^0$  or \htwo\ phases
of star-forming regions 
(Chapter 11 of AGN3).
In this case high-energy photoelectrons, or cosmic rays, can excite and ionize the gas before
undergoing enough elastic collisions to be thermalized.
This is due to the low electron fraction, making it more likely that a suprathermal electron will collide
with an atom or molecule before undergoing thermalizing collisions with electrons.

These suprathermal electrons, sometimes called ``secondary'' or ``knock-on''  electrons,
have been treated in
\citet{1968ApJ...152..971S,1973A&A....25....1B,1979ApJ...234..761S,1991ApJ...375..190X,1985ApJ...298..268S,
1999ApJS..125..237D}.
The  kappa distribution is not used although the idea is similar.
Cloudy has included this physics since its birth in 1978.
The test here is the \emph{ionization fraction}, proportional to $n$(H$^+)/ n$(H),
with suprathermal electrons  being important in neutral regions such as X-ray illuminated 
photodissociation regions (PDRs),
often called XDRs \citep{Maloney1996}.

To summarize this discussion, thermal distributions are established by elastic electron - electron collisions
(Chapter 2, \citealp{1978ppim.book.....S}).
The question whether non-thermal electrons will be important in an H~II region is
really a question of time scales, length scales, and ionization fractions.
Relevant scales in a typical H~II region, the Orion nebula, are discussed next.



\section{The primary mechanism in photoionized nebulae}
\label{sec:primary}

We accept that H~II regions and planetary nebulae are photoionized by starlight. 
The photoionization process will be the main source of high-energy electrons in such nebulae.
As described above, the central question is how quickly, and over what scales, these suprathermal electrons survive
before they become thermal.
First we consider some basic properties of a photoionized cloud.

\begin{figure*}[t]
\includegraphics[scale=0.6]{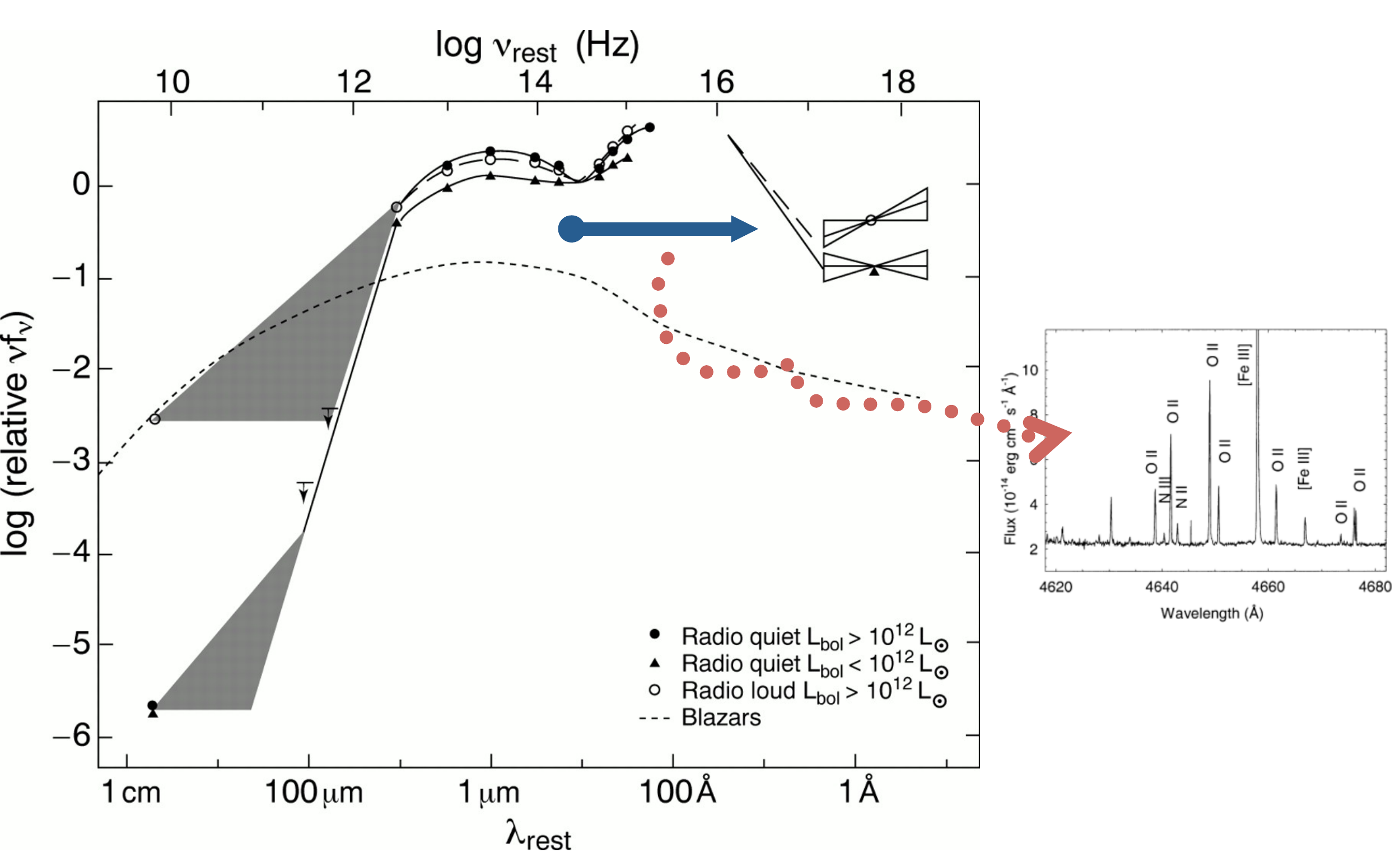}
\caption[The primary mechanism]{The ``primary mechanism'', converting an SED into emission lines.
\label{fig:PrimaryMechanism}}
\end{figure*}

The primary mechanism\footnote{
The term ``primary mechanism'', photoionization by the radiation field of the central object, 
dates back to the 1930's and the original investigations into nebulae.} 
is the process whereby photoionization converts high-energy portions of a stellar SED
into an emission-line spectrum.
It is shown schematically in Figure \ref{fig:PrimaryMechanism}.
In this Figure, taken from AGN3, the  SED
of an active galactic nucleus is shown in the left part of the Figure. 
The rightward pointing arrow shows the energy range of photons which are capable of 
photoionizing hydrogen. 
Ionizing photons are absorbed by atoms 
in a gas cloud, producing  energetic photoelectrons, whose energy is the difference between the ionizing photon and the ionization energy of the atom (AGN3 Chapter 2). 
These energetic electrons collide with thermal electrons and protons to eventually
become thermal.
The free electrons  produce collisionally excited forbidden emission lines through inelastic collisions
with heavy elements,
and briefly become subthermal electrons.
They eventually recombine with an ion, producing recombination emission lines. 
The emission-line spectrum shown in the right of  Figure~\ref{fig:PrimaryMechanism} results. 

\subsection{Details of the Primary Mechanism}

It is commonly assumed that supra or subthermal electrons share their 
energy with surrounding electrons so quickly that they become thermal electrons long 
before exciting one of the forbidden lines of the heavy elements.
\citet{1947ApJ...105..131B} were the first to consider this in detail,
while  \citet{1953PhRv...89..977S} and \citet{1954PhRv...94..511B} go into more details.
This is now textbook material; \citet{Spitzer1962} discusses electron transport at length, while \citet{1978ppim.book.....S} and \citet{2005ppa..book.....K} summarize it more briefly.

\begin{figure}[t]
\includegraphics[scale=0.55]{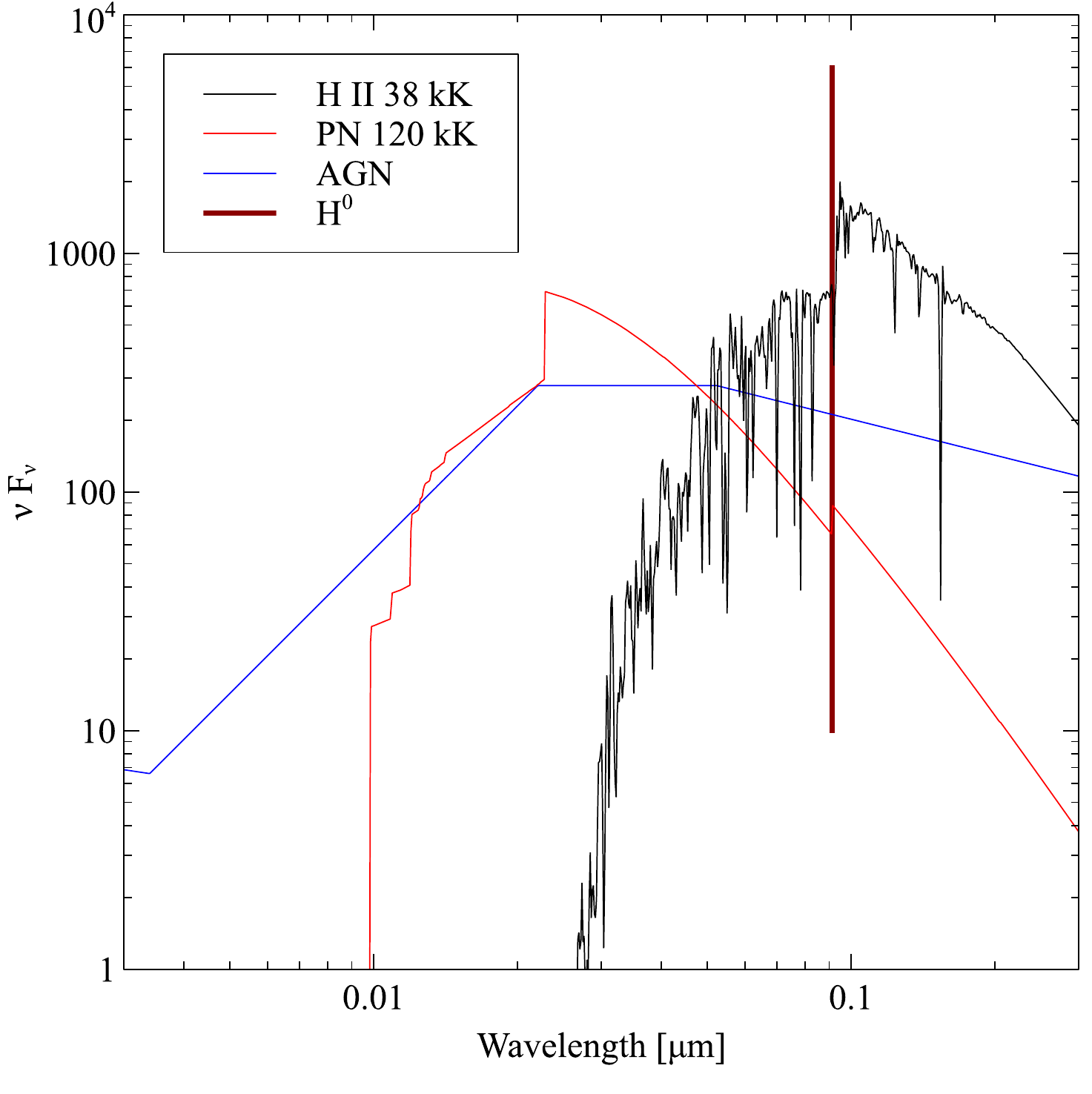}
\caption[Three SEDs]{The SEDs of an O star, a PN central star, and a typical AGN,
as shown.  The vertical line indicates the ionization potential of hydrogen.
\label{fig:sed}}
\end{figure}

The energy of the  photoelectron produced by the photoionization of hydrogen
is central to these timescale questions. Figure \ref{fig:sed} compares the
 SEDs of three different ionizing continua.
The curve with fine structure is the spectrum of an O star 
similar to the ionizing stars in the Orion Nebula. 
The smoothest line extending to the shortest wavelengths and highest energies 
is the SED produced by the central black hole of an AGN. 
The central star of the planetary nebula is the intermediate SED. 
The vertical line in the figure indicates the ionization potential of hydrogen. 
These shapes determine the energy of the photoelectron 
since an ionizing photon produces a photoelectron with an energy equal to the 
difference between its energy and this ionization potential. 
The AGN continuum will produce the most energetic photo-electrons while 
the O star continuum the least.

We can quantify this  by considering an average of the photoelectron energy,
$h ( \nu - \nu_0)$, where $h \nu$ is the ionizing photon energy and  $h \nu_0$
is the ionization potential, 
weighted by the incident photon spectrum ${4 \pi J_{\nu}} / {h\nu}$:
\begin{equation}
\langle  E \rangle = \langle h (\nu - \nu_0 ) \rangle  = \frac{\int_{\nu_0}^{\infty}\frac{4 \pi J_{\nu}}{{h\nu}}\, h(\nu - \nu_0) \, d\nu}
{\int_{\nu_0}^{\infty}\frac{4 \pi J_{\nu}}{{h\nu}}\, d\nu}. 
\end{equation}
Table \ref{tab:PhotoElectronEnergy} gives this mean energy in both
eV and Kelvin units for three different SEDs.  The O star
is the softest of the three continua, producing photoelectrons with a
kinetic energy equivalent to 53~kK, the planetary nebula is
intermediate, and the active galactic nucleus is the hardest SED with
321~kK.

\begin{table}
\centering
\caption{Electron kinetic energies in photoionized nebulae}
\label{tab:PhotoElectronEnergy}
\begin{tabular}{r r r}
\hline
 & $ \langle E \rangle $ & $\langle E\rangle / k $\\
  \hline
  \multicolumn{3}{l}{Photoelectrons:} \\
  H II region SED &  4.54 eV & 52.7 kK \\
  PN SED   & 22.9 eV  & 266  kK \\
  AGN SED  & 27.7 eV  & 321  kK \\
  \hline                        
Thermal electron   & 0.862 eV & 10 kK  \\
\hline
\end{tabular}
\end{table}

For comparison the typical gas kinetic temperature in the surrounding nebulae 
will be about 10k~K. 
Going back to Figure \ref{fig:MaxwellianKappa}, 
a typical photoelectron would be off-scale to the right, 
typically $5 -  30 \sigma$ away from the mean of the Gaussian. 
The central question is, then, whether this high-energy photoelectron can become 
thermal before producing an emission line that we would use as a diagnostic indicator.

\subsection{History of an Electron in the Orion Nebula}
\label{sec:Orion}


We focus on H~II regions because of their role in measuring galactic nucleosynthesis.
We  consider physical processes in the Orion H II region, 
a bright and well studied H~II region with a density of $n($H$) \sim 10^4 \pcc$.
We focus on the model of the bright inner regions developed by \citet{BFM} and
further discussed by \citet{Ferland2001a} and \citet{Ferland2003b}.
We consider the life history of hydrogen  and its   electron
at the midpoint in the H$^+$ region shown in Figure 2 of \citet{Ferland2003b}.
This will establish numbers for the time and distance scale tests.

A neutral hydrogen atom located at this point will survive for roughly five hours before 
an ionizing photon from the Trapezium Cluster causes a photoionization. 
This  generates a photoelectron whose typical energy is given by 
$h\nu - IP \approx 53$~kK. 
The photoelectron is suprathermal  because it has more energy 
than the surrounding thermal electrons, $T \approx 10$~kK. 

The supra-thermal electron remains very energetic for about a second before it 
shares its energy with other free electrons in the gas and becomes thermal. 
The energy exchange occurs through electron - electron collisions, which are among the fastest
collisions in an ionized gas.
They are also perfectly elastic  because little radiation is  produced in a homonuclear
collision, a result of the lack of a dipole moment.
This process is referred to as the thermalization of the supra-thermal electron. 

The electron will remain a thermal free electron for about seven years. 
During this time it may collide with ions, probably O$^{2+}$ or O$^+$ since
oxygen is the third most abundant element and these are the dominant ionization stages.
The thermal electron will excite an internal level of the oxygen ion
and such levels can decay and emit the strong optical lines that are prominent in nebulae. 
Such collisions happen about once a day.
After the collision, the free electron will have lost a great deal of its kinetic energy,
becoming sub-thermal, but regains the energy
following  collisions with other free electrons.
The free electron is rethermalized within about a second. 

After about seven years the free electron will have a near encounter with a proton,
be accelerated and radiate much of its kinetic energy, and recombine forming H$^0$.
Electrons  tend to be captured into highly excited states  
which have lifetimes of about 1\e{-5} s to 1\e{-8} s so the electron  quickly 
falls down to the ground state. 
The electron  remains in the ground state for about five hours before another ionizing photon is 
absorbed and the process starts again. 
This is the primary mechanism in nebulae and summarizes the competing processes
that determine the electron velocity distribution. 

%

\subsection{A question of time and length scales}

There are two tests to check whether an electron will undergo sufficient collisions to become thermal.
The \emph{distance test} checks whether the local photoelectric heating rate changes over 
distances that are smaller
than the electron thermalization mean free path.
The \emph{time scale test} checks whether the photoelectric heating rate changes 
more quickly than the electron thermalization timescale.

First compare the heating scale length with the electron mean free path.  
Heating is by starlight photoionization.  
The mean free path of an H$^0$-ionizing photon is
\begin{equation}
\lambda_{912} = [ n({\rm H}^0) \sigma({\rm H}^0) ]^{-1}  .
\end{equation}
We must compare this with the electron thermalization scale
\begin{equation}
\lambda_{e-e} = [ n_e \sigma_{e-e} ]^{-1}  .
\end{equation}

Consider the midpoint in this nebula, where the H$^0$ fraction is roughly 6\e{-4}. 
The hydrogen density is 1\e{4} cm$^{-3}$ so the H$^0$ density is 6~cm$^{-3}$.  
The hydrogen photoionization cross section near threshold is 
$\sigma($H$^0)$ = 6\e{-18} cm$^{-2}$ so the mean free path 
of a hydrogen ionizing photon is $\lambda_{912} \approx 2.8\times 10^{16}$ cm.  
The heating cannot change over length scales smaller than this, the mean free path of an ionizing photon.  

Next consider the electron mean free path.  
The electron density is 1.1\e{4} cm$^{-3}$ if He is singly ionized.  
The electron - electron collision cross section is 0.8\e{-12} / T(eV)$^2$~cm$^2$ 
(equation 198 of \citealt{2005ppa..book.....K}).  
Take 10 eV for the photoelectron initial energy.  
This is a very high energy for an O star photoelectron (see Table \ref{tab:PhotoElectronEnergy})
but will favor the importance of non-thermal ``kappa'' electrons.  
The electron mean free path is then 
$\lambda_{e-e} \approx 1.1\times 10^{10} \, (T(eV)/10~eV)^2$ cm, 
6.4 dex smaller than the heating scale length.  
The heating is constant over physical scales far larger than the distance between 
thermalizing electron collisions.
This distinguishes  H~II regions from neutral regions of shocks  where
the mean free path can be long.

Next compare the heating timescale and the electron thermalization timescale.  
Assuming photoionization equilibrium and that the ionization / recombination rates are equal, 
the heating rate is $(h\nu - h\nu_o) n_e n_p \alpha_B$ (AGN3).
For the same parameters, 
and a temperature of $T_{e} = 1\e{4}$~K,
the heating rate is then 
10~eV $\times 1.1\e{4}$ 
$\pcc\ \times 1.0\e{4} \pcc\  \times $
2.6\e{-13} cm$^3$ s$^{-1}$ = 2.9\e{-4} eV cm$^{-3}$ s$^{-1}$ 
for these parameters.  
The heat content of the gas is 
3/2 nkT = 1.5 $\times$ 2.2\e{4} $\times$ 8.6\e{-5}$\times$ 1\e{4}   = 2.8\e{4} eV cm$^{-3}$.  
The heating timescale is the ratio, 2.84\e{4}/2.9\e{-4} = 9.7\e{7} s, about three years.  

The time for electrons to approach a Maxwellian is given by \citet{Spitzer1962} 
equation 5-26 for the ``self-collision time'':
\begin{equation}
t_c = \frac{m^{1/2} (3kT)^{3/2}}{8\times 0.714 \pi n e^4 Z^4 \, \log\, \Lambda}
= \frac{11.4 A^{1/2} T^{3/2}}{nZ^4 \, \log\, \Lambda} \textrm{s}.
\end{equation}
Assuming $n_e$ = 1.1\e{4} cm$^{-3}$ and Spitzer's numbers the thermalization timescale is $\sim 1.3$~s.  
The electrons approach a Maxwellian on a timescale 10$^{8}$ times faster than the heating time.  
The heating is constant over times far longer than the time needed to set up a Maxwellian.
This is unlike the solar flare case, where heating rates change very quickly.

These two estimates show that the distance an electron travels
before becoming thermal is 6.4 dex shorter than the scale length heating over which the heating changes, 
and that the electron Maxwellian timescale is nearly 8 dex faster than the heating timescale.  
These establish that the electron velocity distribution has time to closely 
maintain a Maxwellian at the local kinetic temperature.

\section{What if there are non-thermal electrons in the ionized gas?}
\label{sec:WhatIf}

Suppose that a non-photoionization process allows high-energy electrons to exist in the ionized gas.
The electron thermalization cross section
decreases as E$^{-2}$, so higher energy electrons are more difficult to thermalize.
Could these high-energy non-thermal electrons produce forbidden line emission before becoming thermalized?  

\begin{figure}[t]
\includegraphics[scale=0.55]{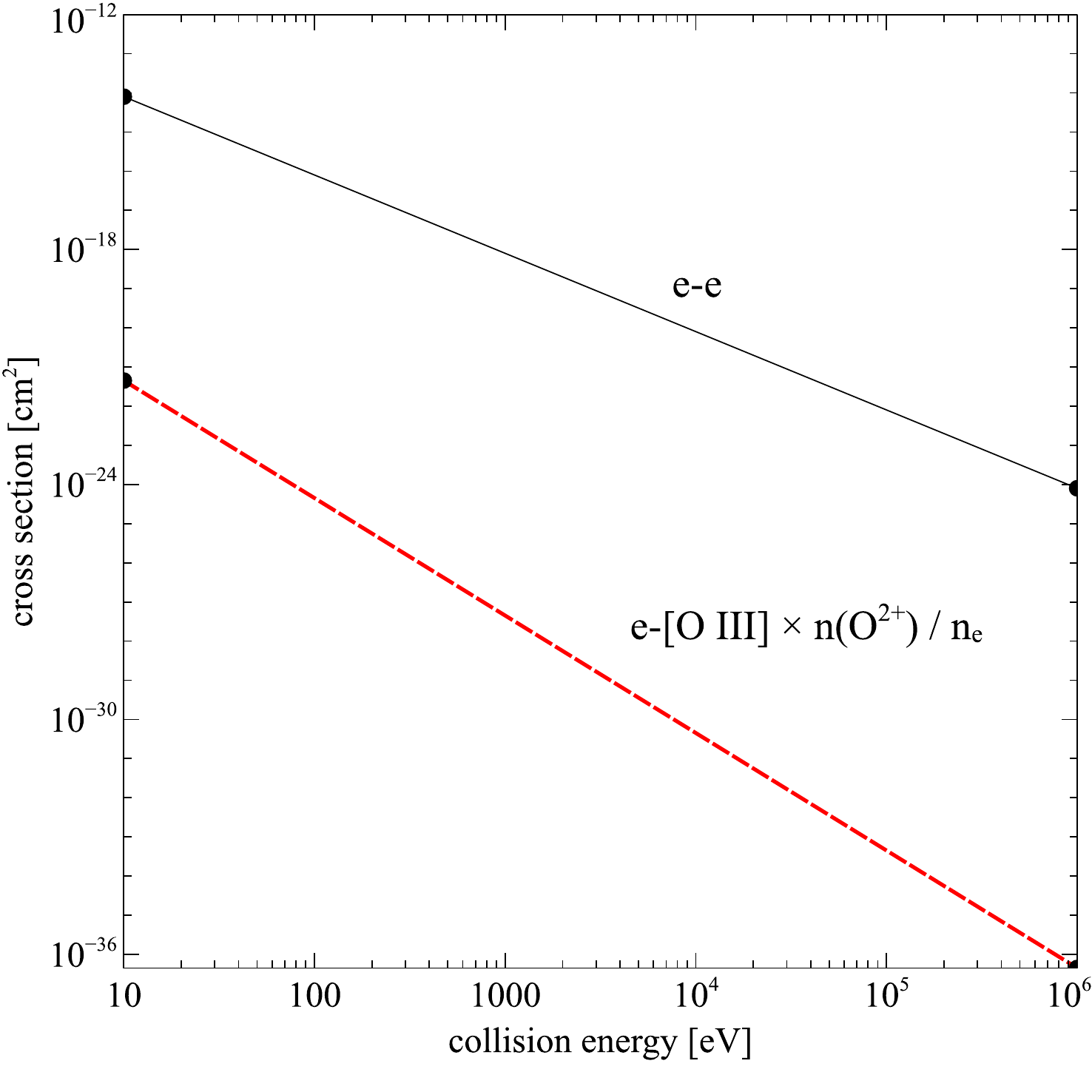}
\caption[Comparing e-e with \[O III\] collision cross sections]{This compares the e-e collision cross section
with the cross section for exciting the O~III green line.}
\label{fig:compareEeO3}
\end{figure}

The non-thermal electron would have to have an inelastic collision with an 
O$^{2+}$ ion and collisionally excite the   500.7 nm [O~III] line 
before it is thermalized by collisions with other electrons for this process to make any sense.  
The cross section for an e-e collision is given by Kulsrud equation 198, 0.8\e{-14}  / T(eV)$^2$ cm$^2$ 
while the cross section for an e - O$^{2+}$ collision that produces the 500.7 nm line is 2\e{-17} cm$^2$
at $\sim 1$~eV. 
This assumed the collision strengths in \citet{1994A&AS..103..273L}.
The cross section for exciting a forbidden line falls off as $E^{-3}$
at high energies \citep{1992A&A...254..436B}.
These are shown in Figure \ref{fig:compareEeO3}.
Assuming a solar O/H, that all O is O$^{2+}$, and $n_e = 1.1 n_{\rm H}$, 
we obtain $n_e / n(\mathrm{O}^{2+}) =$ 2.2\e{3}.
Evaluating the cross sections at 10 keV we find 
$\sigma(e-e) / \sigma({\rm O}^{2+}) \sim 4\times 10^6$ so
that
an energetic electron will have $\sim 9\times 10^{9}$ 
thermalizing collisions with other electrons before it can strike an O$^{2+}$.  
If O$^{2+}$ is many orders above solar, then 
$n(e) \sim n(\mathrm{H}^+) + 2n(\mathrm{O}^{2+})$.
In the presence of only O$^{2+}$, the electron - electron collisions would be about 8\e{6}
times faster than the inelastic collisions with O$^{2+}$.
The result is that an energetic electron will undergo a very large number of thermalizing collisions
long before it can strike an O$^{2+}$ ion.  

Cosmic rays will add kinetic energy to an ionized gas.
They  create a significant population of non-thermal
secondaries in a neutral medium \citep{1968ApJ...152..971S}, but that is not the state of an H~II region,
which is highly ionized.
Given the energy dependencies of the ratio of cross sections given above, very high-energy particles will be even
less likely to directly excite [O~III] lines.
The energy available in cosmic rays is small compared to the energy in starlight in a
typical nebula.

\section{Conclusions}
\label{sec:conclusions}

As discussed in the Introduction, many explanations have been offered for the $t^2$/ADF phenomenon,
and different processes may operate in different objects.
This paper investigates the possibility that a significant population of
non-thermal electrons might be present in ionized regions of nebulae,
and that these disturb the line ratio diagnostics.
Photoionization does produce supra-thermal electrons that are much more energetic than
thermal electrons. 
Cosmic rays or other high-energy particles may also be present.

We do a quantitative evaluation of the time it takes to thermalize such energetic electrons, and the
distances they move in this time.
We have used well established methods to show that the thermalization distances and 
timescales are much smaller than the distance or time in which the heating or temperature
can vary.
These suggest that
supra thermal electrons will have disappeared into the Maxwellian velocity distribution long before they affect the 
collisionally excited forbidden and recombination lines that we use for deriving relative abundances.
We know of no numerical calculations that follow the thermalization of electrons in an ionized gas,
probably because these comparisons suggest that a Maxwellian velocity distribution will result.
These considerations strongly suggest that non-thermal electrons should not be  important 
and cannot account for $t^2$/ADF phenomenon.
Therefore, to explain the observed ADF values in photoionized nebulae, other $t^2$ producing phenomena
must be considered.

Emission line ratios can probe the existence of non-thermal electrons \citep{2013SSRv..178..271B}.
\citet{2013MNRAS.430..599S} considered C~II lines formed by dielectronic recombination
and did not find strong evidence for a kappa distribution.
\citet{2014ApJ...785...91M} consider line ratios, including uncertainties in the atomic data,
and also find no evidence for kappa distributions in nebulae.
Similarly, \citet{2016ApJ...817...68Z} find no evidence of kappa in a sample of H~II regions
and planetary nebulae.
A study of H I emission in Hf 2-2 by \citet{2014MNRAS.440.2581S} finds no evidence of kappa.
Observations support the conclusion that kappa distributions have negligible effect in nebulae.

This work has focused on the Orion H II region, the brightest and best-studied nebula.  A future paper, Henney et al. (in preparation) will extend this analysis to more general cases, including planetary nebulae and extragalactic H II regions.  Quantitative calculations of the deviation from a Maxwellian velocity distribution due to the injection of high-energy electrons will be presented.

\acknowledgments

We thank the referee and Greg Shields for helpful comments.
GJF acknowledges support by NSF (1108928, 1109061, and 1412155), NASA (10-ATP10-0053, 10-ADAP10-0073, NNX12AH73G, and ATP13-0153), and STScI (HST-AR-13245, GO-12560, HST-GO-12309, GO-13310.002-A, and HST-AR-13914).
WJH acknowledges financial support from DGAPA-UNAM through grant PAPIIT-IN11215.
CRO's participation was supported in part by HST program GO 12543.
MP received partial support from CONACyT grant 241732.

\clearpage

\bibliography{LocalBibliography,bibliography2}		

\end{document}